\documentclass[12pt,a4paper]{iopart}

\usepackage{fancyhdr}
\usepackage{epsfig}
\usepackage{graphicx}
\usepackage{verbatim}
\usepackage{latexsym}
\usepackage{colortbl}
\usepackage[numbers,square,sort&compress]{natbib}
\usepackage{hyphenat}
\usepackage[colorlinks]{hyperref}

\begin{document}

\title[Phase diagram of one-, two-, and three-dimensional quantum spin systems]{Phase diagram of one-, two-, and three-dimensional quantum spin systems
              derived from entanglement properties}
\date{\today}
\author{B. Braiorr-Orrs$^1$, M. Weyrauch$^1$, and M. V. Rakov$^2$}
\address{$^1$ Physikalisch-Technische Bundesanstalt, Bundesallee 100, D-38116 Braunschweig, Germany}
\address{$^2$ Kyiv National Taras Shevchenko University, 64/13 Volodymyrska st., Kyiv 01601, Ukraine}
\ead{briiissuurs.braiorr-orrs@ptb.de}

%\affiliation{$^1$ Physikalisch-Technische Bundesanstalt, Bundesallee 100, D-38116 Braunschweig, Germany}
%\affiliation{$^2$ Kyiv National Taras Shevchenko University, 64/13 Volodymyrska st., Kyiv 01601, Ukraine}

\begin{abstract}
  We study the bipartite entanglement per bond to determine characteristic features of the phase diagram of various quantum spin models in different spatial dimensions. The bipartite entanglement is obtained from  a tensor network representation of the ground state wave-function. Three spin-1/2 models (Ising, XY, XXZ, all in a transverse field)  are investigated. Imaginary-time evolution (TEBD in 1D, `simple update' in 2D and 3D) is used to determine the ground states of these models. The phase structure of the models is discussed for all three dimensions.
\end{abstract}
\pacs{64.70.Tg, 03.67.Mn}
\vskip0.5cm
\noindent{\it Keywords\/}: quantum phase transitions, quantum entanglement, tensor networks

%\submitto{\NJP}

\maketitle

%\tableofcontents

\section{Introduction}

It is most intriguing to observe that matter comes in different phases.
Usually, by a change of temperature one may switch between the different phases.
Close to zero temperature quantum fluctuations dominate over thermal fluctuations, and quantum systems  undergo  quantum phase transitions (QPTs) as a result of a change of an external  control parameter~\cite{sachdev:book}.

Discovery of new and exotic quantum phases and critical points, such as topologically ordered phases~\cite{wen:book}, quantum spin liquids~\cite{diep:book}, deconfined quantum critical points~\cite{Senthil05032004, PhysRevB.70.144407}, which are not encompassed by the Landau's theory of phase transitions, recently stirred significant interest into the study of QPTs.
Moreover, it was found that from quantum information theory perspective quantum phases and QPTs can be distinguished and characterized in terms of quantum entanglement~\cite{PhysRevA.66.032110, Osterloh2002}.
Entanglement measures are able to determine critical properties of the systems, in particular the positions of the critical points~\cite{RevModPhys.80.517}.

In the present paper we study how a specific entanglement quantifier, the \textit{bipartite entanglement per bond}~\cite{PhysRevA.81.032304}, can be used for a fast and computationally rather inexpensive determination of quantum critical points and other characteristic features of the phase structure for various one-, two- and three-dimensional quantum spin models.
This quantifier does not require the calculation of expectation values from the states but is obtained directly from the representation of a state. This is the reason why this quantifier comes at low numerical cost. Calculation of expectation values for two and three-dimensional systems is numerically expensive, that is why this quantifier is particularly useful for the analysis in 2D and 3D. The spin-1/2 3D XY model in a transverse field is studied systematically here for the first time.

The entanglement per bond is obtained directly from a tensor network representation of the ground state wave-function.
Tensor networks (TN)~\cite{1751-8121-42-50-504004, doi:10.1080/14789940801912366} provide a modern and promising tool for the numerical investigation of many-body systems.
The basic idea of TN methods is to represent the wave function of a many body quantum system by a network of interconnected tensors. For a recent review see~\cite{orus:review}.
For one-dimensional (1D) systems the best known tensor network states are  matrix product states (MPS)~\cite{doi:10.1080/14789940801912366}. Their direct generalization to 2D and 3D are the projected entangled pair states (PEPS)~\cite{Verstraete:arxiv0407066}. Details about MPS and PEPS can be found in~\cite{SCH10, PhysRevB.81.174411, PhysRevLett.101.090603, PhysRevB.87.085130}.

For our investigation we choose three spin-$1/2$ quantum models with nearest neighbor interactions and analyze them in the thermodynamic limit: the XY model (with quantum Ising model as a special case) and the XXZ model both in a transverse magnetic field. The models are investigated in one- (spin ring), two- (square lattice), and three (cubic lattice) dimensions.
We use the imaginary-time evolution for the determination of the ground states in the TN representation.
The entanglement per bond is calculated directly from the MPS and PEPS representation of the ground states.

The paper is organized as follows. In Sec.~\ref{bipart} we review the definition of the entanglement per bond and comment on its physical meaning. In Sec.~\ref{numerres} we present numerical results along with a detailed discussion. Conclusions are made in Sec.~\ref{conclu}.

\section{Bipartite entanglement per bond}\label{bipart}

In this section we define the bipartite entanglement  per bond $S_{\rm PB}$ in terms of the bond vectors
characterizing the TN. We comment on its physical interpretation and its ability to describe the entanglement properties of the state. Strictly speaking $S_{\rm PB}$ is not an experimentally measurable quantity, since it
depends on a specific representation of a quantum state as a TN.

Ground states in TN representation are obtained either  variationally or via imaginary-time evolution~\cite{CPA:CPA3160070404} based on a Trotter expansion~\cite{trotter} of the evolution operator.
In the present paper we use time-evolving block decimation (TEBD)~\cite{VID2007} to determine ground states of one-dimensional systems and the `simple update' scheme~\cite{PhysRevLett.101.090603} in 2D and 3D. The algorithms we use are well-known, so we do not provide details of their implementation.

The TEBD algorithm naturally leads to a ground state in the canonical form of MPS~\cite{vidal:3,PhysRevLett.93.040502,SCH10}. In this form, besides tensors $A$ at each site one also has  bond vectors $\vec{\lambda}$ at each bond of the tensor network. Within the given canonical MPS the bond states (bond vectors) can be regarded as renormalized bases of the physical degrees of freedom of the many body system (e.g. `effective' spins). Thus bond vectors can be treated as Schmidt coefficients in the wave function decomposition, i.e. they are objects which describe entanglement within the state. Moreover, the reduced density matrix of the state can be expressed approximately via virtual (bond) degrees of freedom, i.e. via the bond vectors. Unfortunately, no {\it exact}  canonical form for PEPS exists, but the `simple update'~\cite{PhysRevLett.101.090603,PhysRevB.86.195137} algorithm for imaginary-time evolution leads to a ground state in {\it approximate} canonical form of PEPS. Thus, bond vectors can be used to obtain the bipartite entanglement per bond in this case as well.
%the concept of so called quasi-canonical PEPS~\cite{PhysRevA.86.022317} was introduced. And the `simple update' algorithm for imaginary-time evolution naturally leads to a ground state in the PEPS representation.

The numerical cost of the TEBD algorithm is $\mathcal{O}(m^3)$ with $m$ the bond dimension. The numerical costs for `simple update' procedure in 2D and 3D are $\mathcal{O}(m^9)$ and $\mathcal{O}(m^{15})$ respectively.
Notice, that the calculation of expectation values from a state given in MPS or PEPS representation requires numerically expensive (especially in 2D and 3D) tensor contraction procedure, but the bipartite entanglement per bond is obtained directly from the wave function tensor network representation, which makes this quantifier so attractive.

The bipartite entanglement per bond $S_{\rm PB}$ is directly defined in terms of the components of the bond vectors $\lambda_i$ which are normalized to satisfy  $\sum_i \lambda_i^2=1$.  Then, $S_{\rm PB}$ is defined as the entanglement entropy~\cite{nielsen:book} or von Neumann entropy
\begin{equation}
S_{\rm PB}= -\sum_i \lambda_i^2 \log_2 \lambda_i^2.
\end{equation}
In practice, we find that the bond vectors connecting to a given site $i$ are approximately equal, which is a consequence of the translational symmetry of the tensor network we implement.  As a consequence, we  average the bond vectors connecting to a site $i$ for the calculation of $S_{\rm PB}$.

Note that the  maximally possible value for $S_{\rm PB}$ (measuring the maximally possible entanglement in the state) is dependent on the  chosen bond dimension $m$ of the MPS or PEPS representation. This means that such a representation may not be able to accurately describe states close to critical points where entanglement may be very
large. Still, the calculated entanglement per bond provides at least semi-quantitative information on the phase structure of the many-body system.

\section{Numerical results and interpretation}\label{numerres}

The present section provides numerical results for the bipartite entanglement per bond  for the Ising, XY and XXZ models in a transverse field in one, two, and three dimensions. Ground states are obtained via imaginary-time evolution algorithms: time-evolving block decimation (TEBD) for modeling ground states of one-dimensional systems and the `simple update' scheme for two- and three-dimensional systems. We choose the spin ring for one-dimensional geometry, the square lattice in 2D and the cubic lattice in 3D always with periodic boundary conditions and translationally invariant. Critical points obtained from this measure dependence are compared to results from previous studies, based on other algorithms.

Imaginary-time steps are proceeded until the convergence of bond vectors is reached. As a check for a good ground state we use the condition of approximate equality of all bond vectors, that means presence of rotational symmetry in the ground state.
Time step size in the imaginary-time evolution is reduced repeatedly during convergence. For TEBD we use MPS bond size up to $m=40$. For the `simple update' algorithm we take bond sizes $m=4$ and $m=2$ in 2D and 3D, correspondingly.

\subsection{Anisotropic XY model}

We first study the anisotropic XY model in a transverse magnetic field,
\begin{equation}
 H^{\rm XY}=-\sum_{\langle i,j\rangle} \left\{ (1+\gamma)  S_i^x   \otimes S_{j}^x + (1-\gamma)  S_i^y   \otimes S_{j}^y \right\}
 +h\sum_i S_i^z,
\end{equation}
where $\langle i,j\rangle$ indicates a summation over nearest neighbors. It includes the Ising model as a special case ($\gamma=1$). We use periodic boundary conditions, and
the spin operators  $S^{\alpha}=\frac{1}{2} \sigma^{\alpha}$ are related to the Pauli matrices $\sigma^{\alpha}$.
The model has two parameters, the anisotropy  $\gamma$ and the magnetic field $h$.

\begin{figure}[h!]
\centering
    \includegraphics[width=0.6\textwidth]{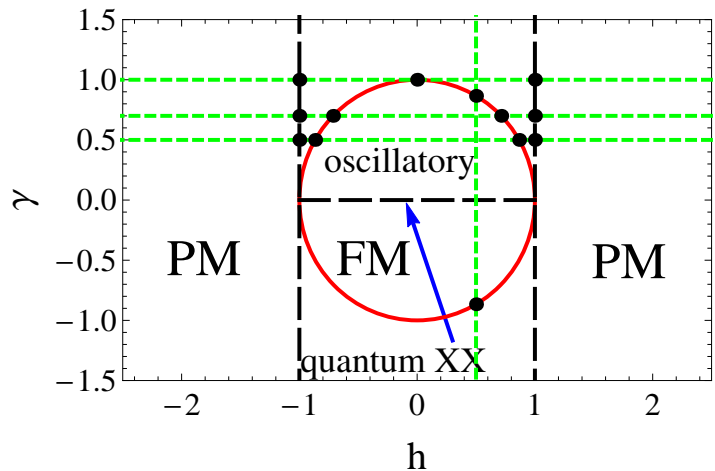}
    \caption{\footnotesize
    Phase diagram of the 1D XY model. Black dashed lines indicate phase separation lines. Green dashed lines $\gamma=1$, $\gamma=0.7$, $\gamma=0.5$ and $h=0.5$ indicate the cases presented in Figs.~\ref{SPB1D2D3D-ising},~\ref{SPB1D2D3D-xy07},~\ref{SPB1D2D3D-xy05}, and~\ref{SPB1D2D3D-xyh1}. The dots mark points where $S_{\rm PB}$ shows special characteristics. (PM: Paramagnetic phase, FM: Ferromagnetic phase).
    }\label{xy1ddiagram}
\end{figure}

The one-dimensional XY model can be solved analytically~\cite{LIE61,sachdev:book,henkel:book}), and as guide for the reader its well-known phase diagram is shown schematically in Figure~\ref{xy1ddiagram}.  There are ferromagnetic and paramagnetic phases which are separated by the black dashed critical lines. On the red circle $h^2+\gamma^2=1$ the model is classical, where we expect entanglement to vanish. It is known that  correlation functions have an oscillatory tail~\cite{PhysRevA.2.1075}
inside this circle. Consequently, this region often is called `oscillatory'. The line $\gamma=0$ separates  two different ferromagnetic phases: the $x$-phase (upper half plain) and the $y$-phase (lower half plain).

The phase structure of the XY model in 2D is similar to the one shown for 1D~\cite{0305-4470-17-14-013, PhysRevA.81.032304}. It is classical for
$(h/2)^2+\gamma^2=1$.
For more details on the phase diagram of 1D and 2D anisotropic XY models see~\cite{1751-8121-43-50-505302,0305-4470-17-14-013}.

The 3D spin-$1/2$ anisotropic XY model in a transverse field has not yet been studied in detail according to our knowledge. Our results suggest that the 3D XY model has a similar phase structure  as its 1D and 2D counterparts, and that it is classical on the circle $(h/3)^2+\gamma^2=1$. However, the bipartite entanglement per bond does not provide enough information to determine the detailed characteristics of the phases. In order to do so, one has  to calculate expectation values of various observables. The entanglement per bond $S_{\rm PB}$ just provides information about the position of critical points as well as lines of vanishing entanglement (`classical' lines).

We now discuss results for $S_{\rm PB}$ along various lines in the phase diagram (indicated as green dashed
in Fig.~\ref{xy1ddiagram}): $\gamma=1$ (Ising model), $\gamma=0.7$ and $\gamma=0.5$ as well as $h=0.5$.

\begin{figure}[h!]
\centering
    \includegraphics[width=0.6\textwidth]{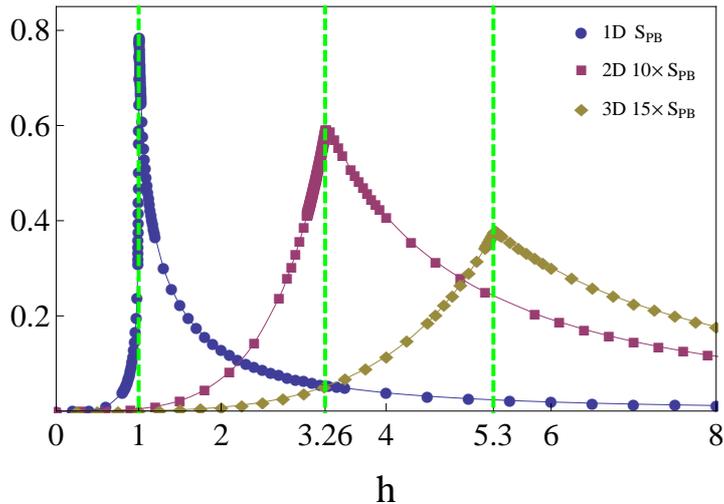}
    \caption{\footnotesize
    Bipartite entanglement per bond for $\gamma=1$ (Ising model) in 1D (chain), 2D (square lattice), 3D (cubic lattice) as a function of the magnetic field $h$.
    2D and 3D results are multiplied by a factor of 10 and 15, respectively.
    Bond dimensions: $m^{\rm 1D}=20$, $m^{\rm 2D}=4$, $m^{\rm 3D}=2$.
    }\label{SPB1D2D3D-ising}
\end{figure}

The results for $\gamma=1$ are shown in Figure~\ref{SPB1D2D3D-ising}.
In 1D our TEBD calculations provide a critical point in very good agreement with the analytical result: $h_{\rm{c}}^{\rm 1D} \approx 1.00$.
The 2D quantum Ising model cannot be solved analytically, and various methods are applied to solve it numerically,
e.g. quantum Monte-Carlo (QMC) methods.
Such calculations find a transition between a ferromagnetic and a paramagnetic phase at a critical point $h_{\rm c}^{\rm 2D}=3.044$~\cite{PhysRevE.66.066110}.
The tensor network implementation applied here produces numerical results significantly faster than QMC calculations,
however, with less precision: it determines a critical point at $h_{\rm{c}}^{\rm 2D} \approx 3.26$. More precise results can be obtained with more elaborate tensor network implementations and larger bond sizes~\cite{orus:review}. However, it is our goal to investigate $S_{\rm PB}$ properties using small numerical cost, i.e. with small virtual bond dimension.

The 3D quantum Ising model cannot be solved analytically as well. The series expansion study in the $T \rightarrow 0$ limit~\cite{0305-4470-27-16-010} predicted the critical point of $h_{\rm{c}}^{\rm 3D} \approx 5.14$. Another study based on `simple update' scheme was performed in~\cite{PhysRevB.87.085130} and the critical point was determined at $h_{\rm{c}}^{\rm 3D} \approx 5.29$ from  the ground state magnetization. The magnetization calculations presented in~\cite{PhysRevB.87.085130} require calculation of expectation values, that is the tensor network contraction procedure. We obtain a result ($h_{\rm{c}}^{\rm 3D} \approx 5.30$) in good agreement with latter one, but without the need to calculate a matrix element
explicitly.

\begin{figure}[h!]
\centering
    \includegraphics[width=0.6\textwidth]{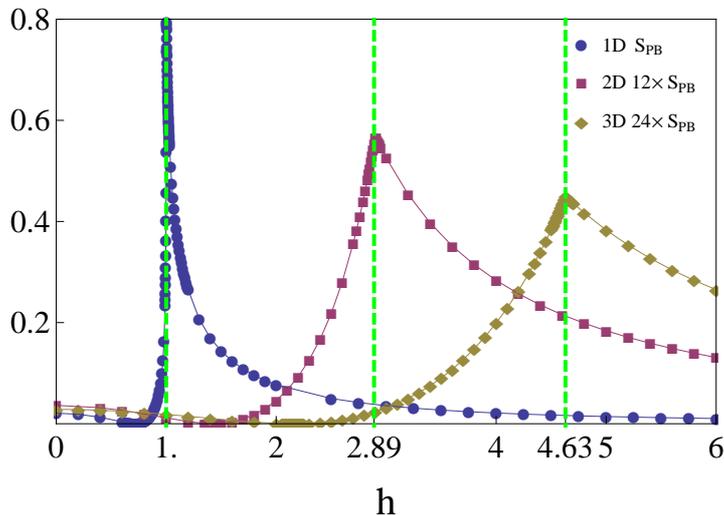}
    \caption{\footnotesize
    Bipartite entanglement per bond for $\gamma=0.7$ in 1D (chain), 2D (square lattice), 3D (cubic lattice)
    as a function of magnetic field $h$.
    2D and 3D results are multiplied by a factor of 12 and 24, respectively.
    Bond dimensions: $m^{\rm 1D}=20$, $m^{\rm 2D}=4$, $m^{\rm 3D}=2$.
    }\label{SPB1D2D3D-xy07}
\end{figure}

\begin{figure}[h!]
\centering
    \includegraphics[width=0.6\textwidth]{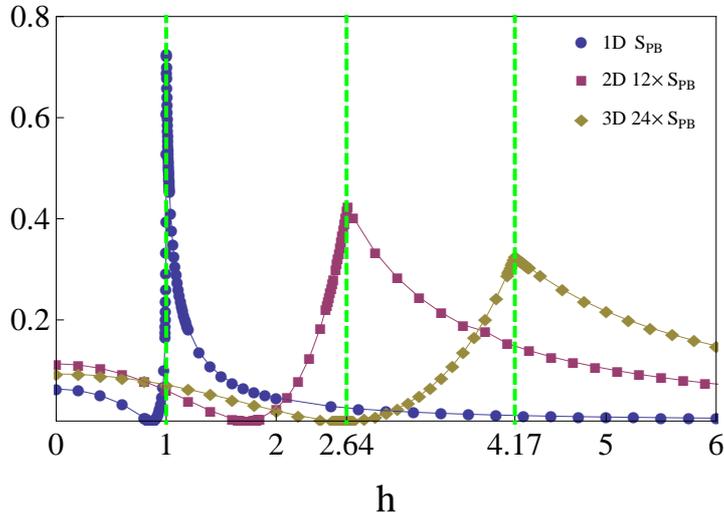}
    \caption{\footnotesize
    Bipartite entanglement per bond for $\gamma=0.5$ in 1D (chain), 2D (square lattice), 3D (cubic lattice)
    as a function of magnetic field $h$.
    2D and 3D results are multiplied by a factor of 12 and 24, respectively.
    Bond dimensions: $m^{\rm 1D}=20$, $m^{\rm 2D}=4$, $m^{\rm 3D}=2$.
    }\label{SPB1D2D3D-xy05}
\end{figure}

Numerical results for $\gamma=0.7$ and $\gamma=0.5$ are shown in the Figures~\ref{SPB1D2D3D-xy07} and~\ref{SPB1D2D3D-xy05}, respectively. $S_{\rm PB}$ has two characteristic features: at the ferromagnetic-to-paramagnetic phase transition  it shows a maximum, and  it is zero at the boundary of the oscillatory region $(h/d)^2+\gamma^2=1$ ($d=1,2,3$). Both points are easily identified in the figures.

In the case $\gamma=0.7$ (Figure~\ref{SPB1D2D3D-xy07}) characteristic features are seen at $h_o^{\rm 1D} \approx 0.71$ and at $h_c^{\rm 1D} \approx 1.00$ for 1D. These results perfectly agree with  theoretical predictions~\cite{PhysRevA.81.032304,1751-8121-43-50-505302}. The 2D result $h_o^{\rm 2D} \approx 1.42$ perfectly agrees with the prediction from $(h/d)^2+\gamma^2=1$, while $h_c^{\rm 2D} \approx 2.89$ differs slightly  from the reference value $h_c^{\rm 2D}=2.72$ obtained from finite size scaling~\cite{0305-4470-17-14-013}. The 3D boundary point $h_o^{\rm 3D} \approx 2.14$  satisfies  $(h/3)^2+\gamma^2=1$, and we obtain the 3D ferromagnetic-to-paramagnetic critical point at $h_c^{\rm 3D} \approx 4.63$.

For $\gamma=0.5$  we obtain in 1D $h_o^{\rm 1D} \approx 0.87$ and $h_c^{\rm 1D} \approx 1.00$ again in perfect agreement with theoretical predictions. Our 2D result $h_o^{\rm 2D} \approx 1.73$  agrees with the theory while $h_c^{\rm 2D} \approx 2.64$ differs slightly (the reference value is $h_c=2.5$~\cite{0305-4470-17-14-013}).  We observe zero entanglement at $h_o^{\rm 3D} \approx 2.60$, which satisfies the expression $(h/3)^2+\gamma^2=1$. The 3D ferromagnetic-to-paramagnetic critical point for $\gamma=0.5$ is determined to be at $h_c^{\rm 3D} \approx 4.17$.

\begin{figure}[h!]
\centering
    \includegraphics[width=0.6\textwidth]{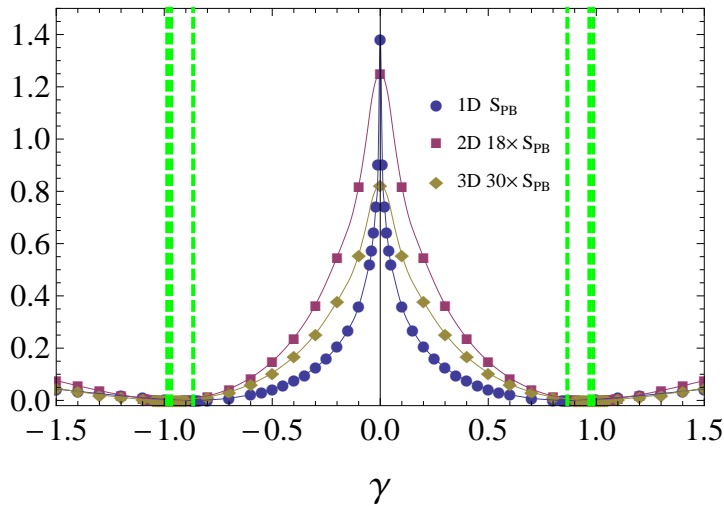}
    \caption{\footnotesize
    Bipartite entanglement per bond for $h=0.5$ in 1D (chain), 2D (square lattice), 3D (cubic lattice)
    as a function of anisotropy parameter $\gamma$.
    2D and 3D results are multiplied by a factors 18 and 30, respectively.
    Bond dimensions: $m^{\rm 1D}=40$, $m^{\rm 2D}=4$, $m^{\rm 3D}=2$.
    }\label{SPB1D2D3D-xyh1}
\end{figure}

\begin{figure}[h!]
\centering
    \includegraphics[width=0.6\textwidth]{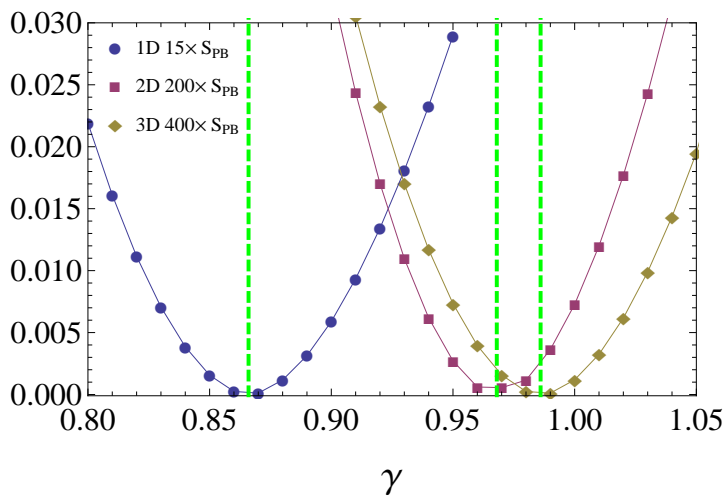}
    \caption{\footnotesize
   Bipartite entanglement per bond for h=0.5 in 1D (chain), 2D (square lattice), 3D (cubic lattice)
    as a function of anisotropy parameter $\gamma$.  Displayed region shows those positive values of $\gamma$, where entanglement vanishes.
    1D, 2D and 3D results are multiplied by a factors 15, 200, and 400, respectively.
    Bond dimensions: $m^{\rm 1D}=40$, $m^{\rm 2D}=4$, $m^{\rm 3D}=2$.
    }\label{SPB1D2D3D-xyh1small}
\end{figure}

Figure~\ref{SPB1D2D3D-xyh1} represents bipartite entanglement per bond for the XY model in all three dimensions as a function of anisotropy parameter $\gamma$ at fixed magnetic field $h=0.5$.
The entanglement measure peaks at $\gamma=0$ indicating the phase transition between two different ferromagnetic phases and vanishes at points which correspond the equation $(h/d)^2+\gamma_o^2=1$.
Figure~\ref{SPB1D2D3D-xyh1small} displays an enlarged view of the region  where entanglement vanishes for positive $\gamma$.
The obtained values for $\gamma^{\rm 1D}\simeq \pm 0.866$, $\gamma^{\rm 2D}\simeq \pm 0.968$, $\gamma^{\rm 3D}\simeq \pm 0.986$ agree to a high precision with the expected theoretical values.

The boundary points of the oscillatory region are obtained numerically up to a very high precision as predicted by the relation $(h/d)^2+\gamma^2=1$. This observation nicely underlines that MPS and PEPS tensor network states are particularly able to model ground states with a low amount of entanglement, that is to model states for which a low bond dimension is enough.

On the other hand, our results for the phase boundaries which are  characterized by a maximum value of $S_{\rm PB}$ differ from the theoretical predictions in 2D presented in Ref.~\cite{0305-4470-17-14-013}.
This is due to the low virtual bond dimension of the PEPS used here, i.e., its impossibility to capture the large amount of entanglement needed to determine the critical point more precisely. Consequently, we would expect that the results for the 3D phase boundaries are even less precise than our 2D results.

From the Figures shown above in this section we conclude that $S_{\rm PB}$ is quite capable of determining important features of the phase diagram for a wide range  of model parameters at relatively small numerical cost. The results for the ferromagnetic-to-paramagnetic phase transition in the 3D XY model, namely $h_c^{\rm 3D} \approx 4.17$ for $\gamma=0.5$ and $h_c^{\rm 3D} \approx 4.63$ for $\gamma=0.7$, were not obtained before to the best of our knowledge.

\subsection{XXZ model}

Next we study the spin-$\frac{1}{2}$ XXZ (anisotropic Heisenberg) model in a transverse magnetic field,
\begin{equation}\label{Ising Hamiltonian}
H^{\rm XXZ}=\sum_{\langle i,j \rangle} \left\{ S_i^x \otimes S_j^x + S_i^y \otimes S_j^y + \Delta \, S_i^z \otimes S_j^z \right\}-\sum_i h \, S_i^z
\end{equation}
as a function of the anisotropy parameter $\Delta$ and the magnetic field $h$.

\begin{figure}[h!]
\centering
    \includegraphics[width=0.6\textwidth]{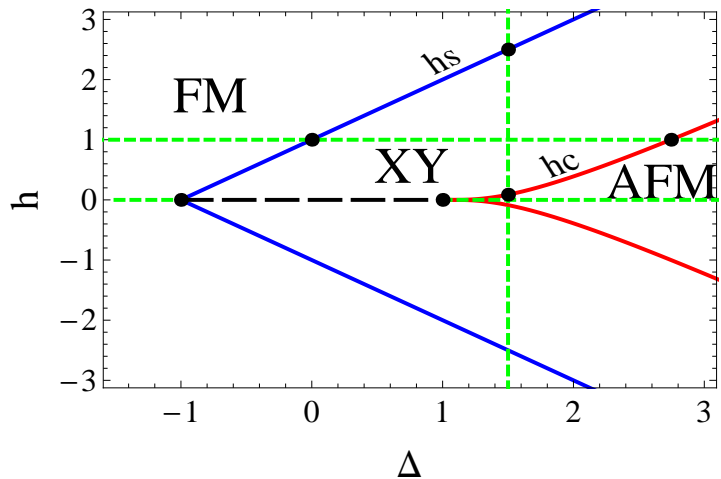}
    \caption{\footnotesize
    Phase diagram of the 1D XXZ model. The line $h_s$ separates the XY phase from the ferromagnetic phase (FM). The curve $h_c$ separates the anti-ferromagnetic (AFM) from the XY phase. The green dashed lines $h=1$ and $\Delta=1.5$ correspond to the cases discussed in detail in the text and in Figures~\ref{SPB1D2D3D-xxz-h1} and \ref{SPB1D2D3D-xxz-delta15}. The black dots indicate the critical points obtained from the entanglement per bond.}\label{xxz1ddiagram}
\end{figure}

The phase structure of the XXZ model was studied extensively in 1D~\cite{PhysRev.150.321, PhysRev.150.327,PhysRev.151.258,PhysRevA.85.052128,  PhysRevB.49.9702, Mikeska01}, 2D~\cite{PhysRevA.85.052128,PhysRevB.65.092402,PhysRevB.64.214411}, and 3D~\cite{PhysRevB.65.092402}.
In all three dimensions the model shows three phases~\cite{PhysRevB.65.092402}: an antiferromagnetic (N\'eel) phase, a XY (spin-flopping) phase and a ferromagnetic  phase. These three phases are separated by two critical lines, and we denote them as $h_c$ and $h_s$.
For 1D the well-known phase diagram can be obtained analytically using the Bethe Ansatz ~\cite{PhysRev.150.321,PhysRev.150.327,PhysRev.151.258}, and it is shown in Figure~\ref{xxz1ddiagram}.
The ferromagnetic phase is separated from spin-flopping phase by the line $h_s=d \, (1+\Delta)$ ($d=1,2,3$ is the dimension of the model)~\cite{PhysRev.135.A640,0953-8984-11-24-311,PhysRevB.65.092402}. In 1D the N\'eel phase is separated from the spin-flopping phase by the curve $h_c$~\cite{PhysRev.151.258,Cloizeaux1966}
\begin{equation}
h_c=
  \frac{\pi \sinh \lambda}{\lambda}\sum_{n=-\infty}^{\infty} {\rm sech} \frac{\pi^2}{2\lambda}(1+2n)
\end{equation}
with  $\lambda= {\rm arccosh} \Delta$.
In particular, at zero magnetic field the ferromagnetic-to-XY critical point is at $\Delta=-1$ and the XY-to-antiferromagnetic critical point at $\Delta=1$.

Similar phase diagrams for the two- and three-dimensional XXZ models, based on quantum Monte Carlo studies, can be found in~\cite{PhysRevB.65.092402}. The most notable difference
to the 1D phase diagram shown here is the fact that the critical lines $h_c$  have a finite derivative with respect to $h$ at $\Delta=1$.

\begin{figure}[h!]
\centering
    \includegraphics[width=0.6\textwidth]{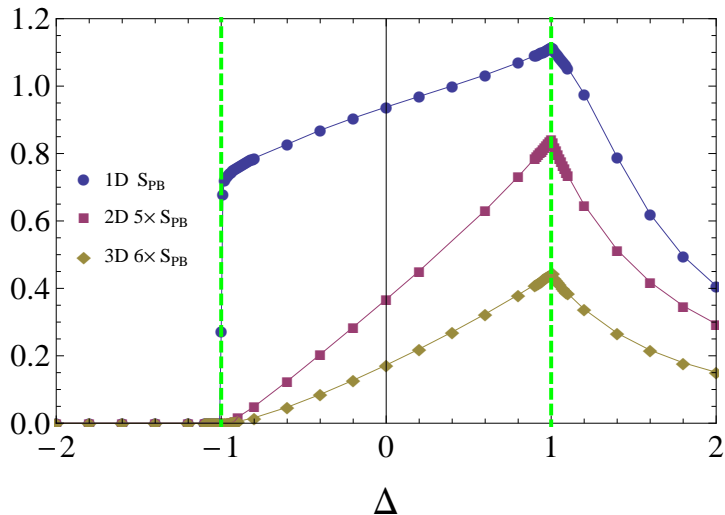}
    \caption{\footnotesize
    Bipartite entanglement per bond for the XXZ model with zero magnetic field in 1D (chain), 2D (square lattice), 3D (cubic lattice)
    as a function of anisotropy parameter $\Delta$.
    2D and 3D results are multiplied a factor of 5 and 6, respectively.
    Bond dimensions: $m^{\rm 1D}=20$, $m^{\rm 2D}=4$, $m^{\rm 3D}=2$.
    }\label{SPB1D2D3D-xxz}
\end{figure}

\begin{figure}[h!]
\centering
    \includegraphics[width=0.6\textwidth]{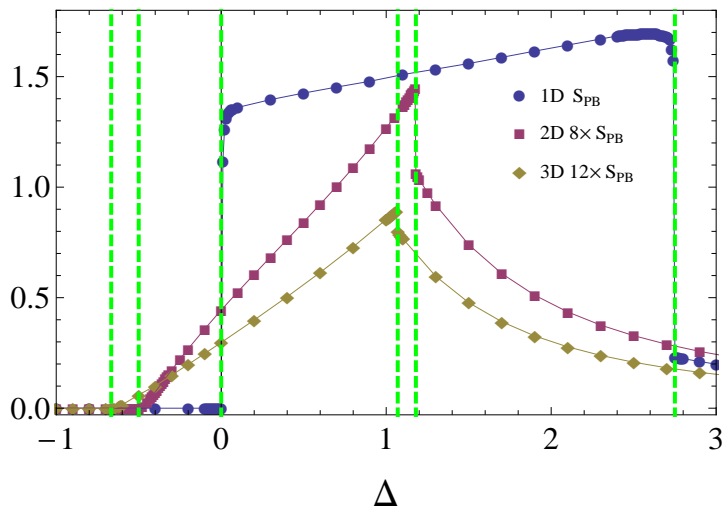}
    \caption{\footnotesize
    Bipartite entanglement per bond for XXZ model with $h=1$ in 1D (chain), 2D (square lattice), 3D (cubic lattice) geometries
    as a function of anisotropy parameter $\Delta$.
    2D and 3D results are multiplied a factor of 8 and 12, respectively.
    Bond dimensions: $m^{\rm 1D}=20$, $m^{\rm 2D}=4$, $m^{\rm 3D}=2$.
    }\label{SPB1D2D3D-xxz-h1}
\end{figure}

\begin{figure}[h!]
\centering
    \includegraphics[width=0.6\textwidth]{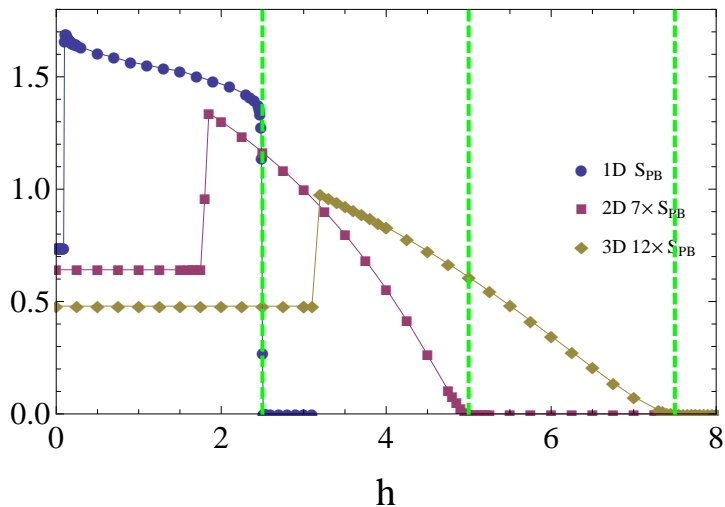}
    \caption{\footnotesize
    Bipartite entanglement per bond for the XXZ model at $\Delta=1.5$ in 1D (chain), 2D (square lattice), 3D (cubic lattice)
    as a function of external magnetic field $h$.
    2D and 3D results are multiplied by factors 7 and 12, respectively.
    Bond dimensions: $m^{\rm 1D}=20$, $m^{\rm 2D}=4$, $m^{\rm 3D}=2$.
    }\label{SPB1D2D3D-xxz-delta15}
\end{figure}

For our numerical studies we choose three cases: $h=0$ (Figure~\ref{SPB1D2D3D-xxz}) and $h=1$  (Figure~\ref{SPB1D2D3D-xxz-h1}) as well as $\Delta=1.5$  (Figure~\ref{SPB1D2D3D-xxz-delta15}). These lines are denoted as green dashed lines in the phase diagram \ref{xxz1ddiagram}.

Figure~\ref{SPB1D2D3D-xxz} presents our results for the bipartite entanglement per bond at zero magnetic field as a function of $\Delta$. From the figure we see that critical points $\Delta=\pm 1$ are obtained correctly. Notice that the critical point $\Delta=1$ is characterized by a maximum of the entanglement, but that at the critical point $\Delta=-1$ the entanglement vanishes. Moreover we observe that at $\Delta=-1$ $S_{\rm PB}$ shows a discontinuity in 1D, while it increases slowly in 2D and 3D.

%This difference of $S_{\rm PB}$ in the XY phase for different dimensions requires comment: the difference stems from the symmetry structure of the ground states. The U(1) symmetry of the XXZ model is predicted to be unbroken in %1D due to the Mervin-Wagner theorem but broken in 2D and 3D~\cite{PhysRevA.68.060301}.  Therefore, one expects to have some types of entanglement (in particular $S_{\rm PB}$) to be constant within the whole XY phase in 1D model. %However, small numerical noise during the TEBD calculations slightly breaks this symmetry also in 1D case~\cite{braiorr2014}, which leads to a slight increase of entanglement within the XY phase.

Figure~\ref{SPB1D2D3D-xxz-h1} shows our results for the bipartite entanglement per bond as a function of $\Delta$ at $h=1$. The 1D critical points are in a very good agreement with exact values~\cite{PhysRev.150.327,PhysRev.151.258}: the ferromagnetic to spin-flopping phase transition occurs at $\Delta_s^{\rm 1D}=0$ and the spin-flopping to antiferromagnetic transition at $\Delta_c^{\rm 1D}=2.75$. The 2D critical points are found at $\Delta_s^{\rm 2D}=-0.5$ and $\Delta_c^{\rm 2D}=1.18$, and the 3D critical points at $\Delta_s^{\rm 3D}=-0.67$ and $\Delta_c^{\rm 3D}=1.07$. The values obtained for $\Delta_s$ in all three dimensions correspond to the relation $h_s=d \, (1+\Delta)$ with dimensions $d=1,2,3$. Our values obtained for $\Delta_c$ for the two- and three-dimensional XXZ model are in a good agreement with quantum Monte Carlo results given in~\cite{PhysRevB.65.092402}.

Figure~\ref{SPB1D2D3D-xxz-delta15} shows results for $S_{\rm PB}$ as a function of the magnetic field $h$ at  $\Delta=1.5$.
The XXZ model at $\Delta=1.5$ was studied analytically in 1D~\cite{PhysRev.150.327,PhysRev.151.258} and was investigated numerically in 2D and 3D~\cite{PhysRevA.81.032304,PhysRevB.65.092402}. These studies predict two critical points for each dimensionality. From equation $h_s=d \, (1+\Delta)$ one obtains $h_s^{\rm 1D}= 2.5$, $h_s^{\rm 2D}= 5.0$, $h_s^{\rm 3D}= 7.5$, and our results agree with these values. For the one-dimensional model the exact solution gives $h_c^{\rm 1D} \simeq 0.0866$~(\cite{PhysRev.150.327}), and our result $h_c^{\rm 1D} \simeq 0.09$ is very close. Our 2D and 3D results are $h_c^{\rm 2D} \approx 1.8$ and $h_c^{\rm 3D}\approx 3.2$ are in a good agreement with previous quantum Monte Carlo studies~\cite{PhysRevB.65.092402}.

From  Figures~\ref{SPB1D2D3D-xxz},~\ref{SPB1D2D3D-xxz-h1}, and~\ref{SPB1D2D3D-xxz-delta15} we conclude that $S_{\rm PB}$ is capable of providing quite some insight into the phase structure of a quantum spin model. One not only obtains critical points, but also points or lines where the system behaves classically. The variation of $S_{\rm PB}$ in between such lines provides insight how the entanglement of the various ground states changes as a function of the control parameters.
Of course, for a more detailed investigation one then would need to calculate relevant expectation values in order to obtain physical properties in interesting regions. The identification of such regions is easily guided
by the $S_{\rm PB}$, which is obtained at a rather low numerical cost.

From Figures~\ref{SPB1D2D3D-xxz},~\ref{SPB1D2D3D-xxz-h1}, and~\ref{SPB1D2D3D-xxz-delta15} we observe interesting differences for $S_{\rm PB}$ at the critical points in different spatial dimensions. E.g., at the transition from the ferromagnetic to the spin-flopping phase $S_{\rm PB}$ shows a jump in 1D, for 2D and 3D it shows a cusp. The numerical results indicate that the entanglement structure of the ground state of the XXZ model in different spatial dimensions differs not only quantitatively but also qualitatively. Understanding the reasons for these differences requires further analysis which is beyond the bipartite entanglement per bond quantifier.

\section{Conclusions}\label{conclu}

In the present paper the bipartite entanglement per bond  $S_{\rm PB}$ was calculated for the Ising, anisotropic XY and XXZ models in a transverse magnetic field and in one, two, and three dimensions for various sets of control parameters.
All obtained results are in very good agreement with previous studies. They show that bipartite entanglement per bond is an efficient, easy-to-calculate and relatively precise tool for the determination of the some characteristics  of the phase diagram of quantum models.
It conveniently uses the tensor network representation of a state.

The three-dimensional anisotropic XY model with transverse field on a cubic lattice was studied numerically for the first time. The obtained boundary points of the oscillatory region satisfy the relation $(h_o/3)^2+\gamma^2=1$.

We observe that the bipartite entanglement per bond determines  characteristic features of the phase diagram better if states with a low amount of entanglement are involved. For our 2D and 3D calculations we took rather low virtual bond sizes and therefore our PEPS can only model states with rather low entanglement. Moreover, if more detailed
questions about the phase diagram need to be answered for specific regions (e.g. the magnitude of the magnetization), then calculations of matrix elements cannot be avoided.

The present paper was geared towards a semi-quantitative analysis of the bipartite entanglement per bond. An interesting and promising task is the quantitative investigation which requires implementation of the symmetries of the ground states into the imaginary-time evolution algorithms. That would enable higher virtual bond dimensions for MPS and PEPS and the calculation of the entanglement spectrum of the state.

Besides the  geometries studied in the present paper (spin ring, square lattice, cubic lattice), the bipartite entanglement can also be used to characterize properties of  quantum models on hyperbolic lattices~\cite{PhysRevE.86.021105}, after modification of the `simple update' algorithm for the ground-state calculation. Moreover, the present study should be extended to spin-1 systems.

%\section*{References}
%\InputIfFileExists{entanglement per bond last.bbl}

{\small
\bibliographystyle{unsrtNoTitles}
\bibliography{spins,refs}
}

\end{document}